\documentclass[twocolumn,prl]{revtex4}
\usepackage{graphicx}
\usepackage{amsmath}
\usepackage{amssymb}
\usepackage{bm}
\usepackage{float}
\usepackage{subfigure}
\usepackage{amsfonts}
\usepackage{setspace}
\usepackage[usenames]{color} 
\usepackage{thumbpdf}
\newcommand{\be}{\begin{equation}} \newcommand{\ed}{\end{displaymath}}
\newcommand{\bd}{\begin{displaymath}} \newcommand{\ee}{\end{equation}}
\newcommand{\bea}{\begin{eqnarray}} \newcommand{\eea}{\end{eqnarray}}
\newcommand{\ba}{\begin{array}} \newcommand{\ea}{\end{array}}

\begin{document} 
\title{Chameleon Effects on Small Scale Structure and the Baryonic Jeans Mass}
\author{Katherine Jones-Smith}

\affiliation{Physics Department and McDonnell Center for the Space Sciences, 
Washington University, St Louis, MO 63130, USA}
\begin{abstract}
In the framework of Newtonian cosmology or general relativity it is simple to derive a mass scale below which collapsed structures are relatively devoid of baryons. We examine how the inclusion of a chameleon scalar field affects this baryonic Jeans mass, bearing in mind both the canonical case of a gravitational-strength coupling between the scalar field and matter, as well as the strong coupling regime wherein the coupling is very large. We find that baryon effects persist down to smaller scales in a chameleon theory than they do in ordinary general relativity, especially in the case of strong coupling. Several potentially observable consequences of this are identified.  
\end{abstract}

\maketitle

Modified theories of gravity often invoke a screening mechanism to remain consistent with precision tests of gravity on the
terrestrial and solar system scale \cite{khouryarxiv, carroll, capo, odin, husawicki, starob}. The chameleon scalar field \cite{kwprl, kwprd} offers one such mechanism; in this family of models a scalar field $\phi$ affects matter through a conformal coupling $A^2(\phi)$. 
The Einstein-Hilbert action
is unchanged from ordinary GR but the matter fields are assumed to be coupled to a Jordan frame metric $\tilde{g}_{\mu\nu}=A^2(\phi)g_{\mu \nu}$. 
Assuming a Minkowski metric, negligible time evolution of $\phi$ and non-relativistic matter of density $\rho$, this coupling has the effect of introducing a density dependence in the effective potential 
which governs the evolution of the field: 
\be
\nabla^2 \phi = {\partial^2 V_{eff}} /{\partial \phi^2}
\ee
where
\be
V_{eff}=V(\phi)+\rho A(\phi),
\ee

The presence of this term in the potential means that for suitably chosen  $V(\phi)$ and $A(\phi)$ the ambient matter density dictates the mass of the field $m^2=\partial^2 V_{eff}/\partial \phi^2$.  In high density environments like the Earth the mass becomes very large relative to low density environments like the cosmos where the field could be so light as to play the role of dark energy \cite{kwprl, kwprd}. This ability to `blend in' with its surroundings earns the field its name. 

The purpose of this paper is to connect 
several recent findings regarding the chameleon model.
The original chameleon model assumed $A(\phi)=e^{\beta \phi/M_{Pl}}$, where $\beta$ is the coupling constant and is taken to be order unity, corresponding to a gravitational-strength coupling between matter and the chameleon field. This model demonstrates the surprising result that a scalar field can couple to matter non-negligibly and yet evade terrestrial constraints on equivalence principle violation and deviations from the gravitational inverse-square law \cite{kwprl, kwprd}.  However, Mota and Shaw later
demonstrated \cite{motashaw} that in fact the chameleon could be much more strongly coupled ($\beta \gg 1$) and still evade terrestrial detection. 
 
On the other hand, Brax {\em et al.} have studied
the evolution of dark matter density perturbations 
in chameleon cosmology\cite{brax}, and found that on sufficiently small scales the perturbations
grow with an exponent that depends linearly on $\beta$. Even for the case of $\beta \sim 1$ small scales demonstrate dramatic enhancement within the chameleon model. 
Thus we are motivated to ask whether the strong coupling regime would produce astrophysical signatures in the realm of small scale structure.

The analysis of Brax {\em et al.} does not include the effect of baryon pressure on the evolution of cold dark matter (CDM)
perturbations. It is well-known that in ordinary GR baryon
perturbations follow the CDM perturbations on large scales \cite{peebles}; 
due to the acoustic pressure of baryons
there is a Jeans scale below which
collapsed structures have negligible baryon content.
To illustrate the kinds of effects the chameleon might
have on small scale structure, here we study the same problem in the context of 
chameleon cosmology by generalizing the analysis of Brax {\em et al.} to
include baryons and acoustic pressure.
 
First we review the baryonic Jeans calculation within general relativity, {\em i.e.}
the growth of  linear perturbations in CDM and baryonic matter density, 
from decoupling at redshift $z_{dec} \sim 1100$ to the redshift at which baryons and radiation are
still at the same temperature, $z=z_b\sim 150$. 
The presentation here closely follows Section 8.3 of Weinberg \cite{weinberg}. First order 
density
perturbations of CDM 
and baryons ($\delta \rho_D$
 and $\delta \rho_B$ respectively), 
to their otherwise uniform backgrounds 
$\overline{\rho}_D$ and 
$\overline{\rho}_B$  
obey identical continuity equations 
 \be 
\frac{d \delta\rho_D}{dt}+3H\delta\rho_D- a^{-1}\overline{\rho}_D {\bf q}^2 \delta u_D = 0 
\label{rhodcont}
\ee  
\be 
\frac{d \delta\rho_B}{dt}+3H\delta\rho_B- a^{-1}\overline{\rho}_B {\bf q}^2 \delta u_B = 0,
\label{rhobcont}
\ee  
at co-moving wave-vector ${\mathbf q}$. 
However, the Euler equation for each type of matter is different, reflecting the fact that baryons experience pressure. The baryonic pressure is reflected in the speed of sound, denoted $v_s^2 = \partial p_B/\partial \rho_B$. Thus, the Euler equations are 
\be 
 \frac{d \delta u_D}{dt} + H\delta u_D=\frac{4\pi G a}{q^2}\left(\delta \rho_D + \delta \rho_B\right),
\label{rhodeuler}
 \ee 
\be 
 \frac{d \delta u_B}{dt} + H\delta u_B=\frac{4\pi G a}{q^2}\left(\delta \rho_D + \delta \rho_B\right)-\frac{v_s^2}{a\overline{\rho}_B}\delta \rho_B.
 \label{rhobeuler}
 \ee 
Note that we can eliminate the velocity potentials 
$\delta u_D$ and $\delta u_B$
via Eqs (\ref{rhodcont}) and (\ref{rhobcont}). Writing $\delta_D \equiv \delta \rho_D/\overline{\rho}_D$ and $\delta_B \equiv \delta \rho_B/\overline{\rho}_B$, and using the fact that both $\overline{\rho}_D$ and $\overline{\rho}_B$ fall off as $a^{-3}$ or $t^{-2}$ in this era, we can write the Euler equations (\ref{rhodeuler}) and (\ref{rhobeuler}) as 
\be 
\ddot{\delta}_D +\frac{4}{3}\dot{\delta}_D=\frac{2}{3t^2}\left(f\delta_B + (1-f)\delta_D\right)
\label{ddoubledot}
\ee
and 
\be 
\ddot{\delta}_B +\frac{4}{3}\dot{\delta}_B=-\frac{2\alpha}{3t^2}\delta_B+  \frac{2}{3t^2}\left(f\delta_B + (1-f)\delta_D\right)
\label{bdoubledot}
\ee
where 
\be
\alpha \equiv \frac{3{\bf q}^2v_s^2 t^2}{2a^2}=\frac{v_s^2{\bf q}^2}{4\pi G\; \overline{\rho}_M a^2}, \;\;\; f\equiv \frac{\overline{\rho}_B}{\overline{\rho}_M}= \frac{\Omega_B}{\Omega_M}
\label{defns}
\ee
and $\overline{\rho}_M \equiv \overline{\rho}_D+\overline{\rho}_B$.

Fo the epoch we are considering, roughly  $1100 < z < 150$, the temperature $T$ of baryonic matter was the same as radiation
and $v_s^2 \propto T \propto a^{-1}$, so $\alpha$ was constant. Thus we can seek power law solutions to Eqs. (\ref{ddoubledot}) and (\ref{bdoubledot}).  If we write
\be
\delta_D \propto t^\nu,\;\;\;\; \delta_B = \xi \delta_D,
\label{ansatz}
\ee
Note that the two parameters $\nu$ and $\xi$
encompass the effects we wish to examine when we generalize to the chameleon case: the growth of the perturbations and the relative strength of the baryon and CDM perturbations. We assume they are independent of time but not wave-vector {\bf q}. With these assumptions Eqs. (\ref{ddoubledot}) and (\ref{bdoubledot}) become 
\be
\nu^2+\frac{\nu}{3}=\frac{2}{3}(f\xi+(1-f)),
\ee
\be
\nu^2+\frac{\nu}{3}+\frac{2\alpha}{3}=\frac{2}{3}(f+(1-f)/\xi). 
\ee
Eliminating $\xi$ yields a quartic equation for $\nu$; Weinberg makes the approximation that $f=0$ and obtains the relevant (growing) solution 
\be 
\nu=2/3, \;\;\; \xi=\frac{1}{1+\alpha}.
\label{bigstevenu}
\ee

It is convenient to write $\alpha = q^2/q_J^2$ where 
\be
q_J^2 = \frac{9}{10} \Omega_M H_0^2 \frac{\mu m_N}{k_B T_{\gamma 0}}.
\label{eq:qj}
\ee
Here $m_N$ is the mass of a hydrogen atom and 
$\mu = 1.22$ takes in to account that matter was a mixture of hydrogen and helium atoms. $H_0$ is the
present value of the Hubble constant and $T_{\gamma 0}$ is the present temperature of the cosmic
microwave background. 
In going from the expression for $\alpha$ in eq (\ref{defns}) to the expression above
we have made use of the formula for the speed of sound in a gas 
$v_s^2 = 5 k_B T/3 \mu m_N$. Plugging in these values we
find that $q_J = 384 h $ (Mpc)$^{-1}$ corresponding to a length scale 
$\lambda_J = 2 \pi/q_J = 16/h$ kpc. The Jeans mass
$M_J$ is then defined as the amount of matter in a sphere of radius $\lambda_J$ at the
present epoch. Numerically $M_J \approx 6 \times 10^5 M_{\odot}$. 

Physically, $\xi > 1$ corresponds to baryon enrichment and $\xi < 1$ to baryon depletion. 
Eq (\ref{bigstevenu}) reveals that $\xi \approx 1$ for small wave-vectors $q \ll q_J$ and $\xi \rightarrow 0$
large wave-vectors $q \gg q_J$. In other words on large scales baryon perturbations follow dark matter 
perturbations but on small length scales there is significant baryon depletion. Equivalently one can say that collapsed structures
of mass less than $M_J$ have significant baryon depletion.

It is straightforward to extend this analysis to include a chameleon scalar field. 
 Brax {\em et al.} showed that on sufficiently small scales the effect of the chameleon 
on the evolution of perturbations is simply to enhance $G$ by a factor of $(1 + 2 \beta^2)$.
On such scales Eqs. (\ref{ddoubledot}) and (\ref{bdoubledot}) therefore become
\be 
\ddot{\delta}_D +\frac{4}{3}\dot{\delta}_D=\frac{2}{3t^2}(1+2\beta^2)\left(f\delta_B + (1-f)\delta_D\right)
\label{ddoublecham}
\ee
and 
\be 
\ddot{\delta}_B +\frac{4}{3}\dot{\delta}_B=-\frac{2\alpha}{3t^2}(1+2\beta^2)\delta_B+  \frac{2}{3t^2}\left(f\delta_B + (1-f)\delta_D\right).
\label{bdoublecham}
\ee
Employing the same power law ansatz, we have
\be
\nu^2 +\frac{\nu}{3}=\frac{2}{3}(1+2\beta^2)(f\xi+(1-f)),
\ee
\be
\nu^2+\frac{\nu}{3}+\frac{2\alpha}{3}=\frac{2}{3}(1+2\beta^2)(f+(1-f)/\xi). 
\ee
We need not resort to the putative smallness of $f$ to obtain a solution; the quartic equation for $\nu$ obtained by eliminating $\xi$ can be solved exactly. The solutions are 
 \be
 \nu_{+\pm}=\frac{1}{6}(-1 + \sqrt{1+36n_{\pm}})
 \ee
and
\be
 \nu_{-\pm}=\frac{1}{6}(-1 - \sqrt{1+36n_{\pm}})
 \ee
where 
\begin{align}
n_{\pm}&= \frac{-1}{3}(\alpha- (1+2\beta^2))\pm \gamma \\
\gamma&=\sqrt{(\alpha-(1+2\beta^2))^2 +4\alpha(1+2\beta^2)(1-f)}
\end{align}
 The only non-decaying solution is $\nu_{++}$. From this we determine 
 \begin{align}
 \xi&=\frac{2(1+2\beta^2)(1-f)}{\alpha+(1+2\beta^2)(1-2f)+\gamma}.
 \end{align}
We have left $f$ unspecified, which makes it easy to check that this reduces to the solution Eq.(\ref{bigstevenu}) for $f=\beta=0$. The baryon fraction to good approximation is $f=1/6$; for strong enough coupling such that $1+2\beta^2\approx 2\beta^2$,  this gives 
\begin{align}
\nonumber
\nu_{sc}&=\frac{-1}{6}+\frac{1}{6}\left \{1+36(-\frac{1}{3}(\alpha-2\beta^2) \right. \\
&+\left. \sqrt{(\alpha-2\beta^2)^2+\frac{20}{3}\alpha \beta^2}\right \}^{1/2}
\end{align}
where we have written the terms out explicitly to demonstrate the dependence of  $\nu$ on $\beta$. But for present purposes we are more interested in the relative strength of the baryon perturbation; defining $y\equiv q/\beta q_J$, we have 
\be
\xi_{sc}=\frac{20/6}{y+\frac{4}{3}+\sqrt{(y+2)^2-\frac{4}{3}y} }
\label{xisc}.
\ee

Note that the above approximation is not so bad even for $\beta =1$, and the overall behavior of $\xi$ is the same, but more specifically for this case we have
\be
\xi_1=\frac{5}{y^2+2+\sqrt{(y^2+3)^2-2y^2} }.
\label{xi1}
\ee
We plot the behavior of $\xi$ for ordinary CDM and the two chameleon models in Figure 1. Clearly the attenuation of baryon presence with wave number is much less rapid in the chameleon case of $\beta=10$ than for $\beta=1$ or ordinary GR. \\
\begin{figure}
\includegraphics[scale=.7]{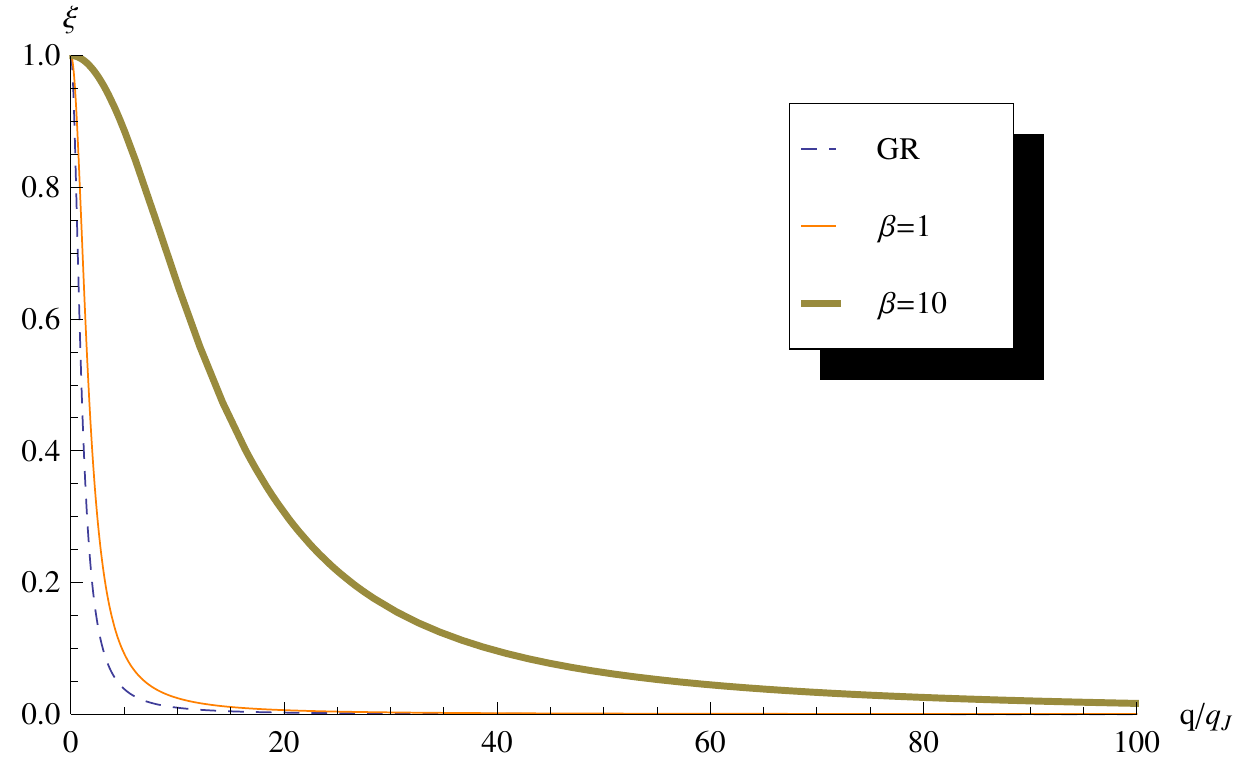}
\caption{Relative strength of baryon to dark matter perturbation as a function of wave-vector for ordinary GR/CDM (dashed) , and a chameleon model with $\beta=1$ (orange) and $\beta=10$ (green).  The $\beta=10$ case differs significantly from CDM case; baryon perturbations persists at smaller structures in a chameleon theory with $\beta > 1$ relative to normal CDM.  As discussed in the text, the chameleon only affects modes above a characteristic scale $q_{cham}$; modes much greater than this threshold will be influenced by the chameleon throughout their evolution and modes much below this scale will evolve unaffected by the chameleon. For modes between these two regimes the chameleon kicks in at some point during their evolution. Generally speaking $q_{cham}$ increases as $\beta$ increases, 
for the fiducial model in Eq.(\ref{vcanon}), modes with 
$ q > q_J \beta^{5/2} (70\; {\rm eV}/M)^{5/4}$
will undergo chameleon-influenced evolution throughout, and those with 
$ q < q_J \beta^{5/2} (70\; {\rm eV}/M)^{5/4} (z_b/z_{dec})^{5/4}$
will not be affected by the chameleon for the range of redshifts considered here. For the potential in Eq.(\ref{vcanon}) it is possible then for a strongly coupled chameleon to evade astrophysical constraints by affecting 
only unobservably small
scales. 
But for the alternative parametrization given in Eq.(\ref{twopa}), $M_2$ replaces $M$ in the threshold wave-vector, and 
sub-galactic scales and scales of order $q_J$ 
would be strongly affected. }
 \label{plot}
\end{figure}
\\
\begin{center}
\noindent{\em Discussion}\\
\end{center}
\noindent What are the implications of the baryon persistence demonstrated in Fig. 1? The index $\nu$ represents the exponent to which linear perturbations in the matter fluid grow with time. Since it acquires a linear dependence on $\beta$ in chameleon models, things quickly get out of control in the strong coupling regime; for example for $\beta=10$,  $\nu(q=100q_J) \approx 80$. Clearly this regime needs to be handled more carefully than the linear treatment, nonetheless several observations can be reliably inferred from the above analysis. 

The chameleon has a characteristic scale $k_{cham}$at which it acts; this is a function of redshift as well as the particulars of the potential $V(\phi)$. Modes much greater than the characteristic scale $q_{cham}$ are the most significantly affected; but as is relevant to the strong coupling regime we note that as  $\beta$ increases, $q_{cham}$ also increases. So generally speaking the strongly coupled chameleon could evade constraints in yet another way: if the scale on which it acts always above observable astrophysical scales. 

But it is easy to construct potentials in which this is not the case. Recall the fiducial potential has one parameter $M$ which has units of mass
\be
V(\phi)=M^4\exp (\frac{M}{\phi})^n, 
\label{vcanon}
\ee 
For $\phi \gg M$ this reduces to an inverse power law
$V(\phi) \approx M^{4+n}/\phi^n$. 
It turns out that the upper bound on M that can derived from terrestrial experiments is the same energy scale as is relevant to dark energy, $M \lesssim 10^{-3}$eV \cite{kwprd}.  A two parameter generalization of this, 
\be
V(\phi)=M_1^4\exp (\frac{M_2}{\phi})^n, 
\label{twopa}
\ee 
has all of the same appealing features as Eq.(\ref{vcanon}). If we assume $M_2 \ll \phi_{\min} \ll M_{Pl}$, then 
\be
V_{eff}(\phi_{min}) \sim M_1^4 
\ee
and if we choose $M_1 \sim 10^{-3}$ eV this is consistent with the dark energy scale. It is straightforward to show that the slow roll conditions and tracker solution behavior are met as well.  While $M_1$ could supply the requisite dark energy scale,  $M_2$ could operate on scales such that it affects the growth of sub-galactic scales throughout their evolution.  
In order to do so we would need  
$\beta q_J / a \gtrsim m$
for roughly the epoch of $z_{dec}$ to $z_b$.  For $n=1$, these considerations amount to requiring  
$ \beta^{1/5} M_2 \gtrsim 70$ eV.
  
Such a model could have interesting astrophysical effects,  for example in the context of the so-called missing satellite problem. This phrase refers to the dearth of observed satellite galaxies in the vicinity of Milky Way sized galaxies compared to what is predicted by numerical simulations and analytic estimates (see \cite{bullock} for an introduction).  It is a rather complicated situation and much of the astrophysics is poorly understood and/or constrained. For example is not clear whether the satellites are just too faint to be detected at present, or if they are actually absent.  But surely, a persistence of baryons at sub-galactic scales stands to worsen the problem, and as Fig. 1 indicates chameleon models seem to demonstrate such behavior at strong coupling.  So it is conceivable that the strong coupling regime could be constrained by existing astrophysical observations, such as the population of dwarf spheroidals.  This is the topic of future work.

\begin{center} 
{\em Acknowledgments}\\
\end{center}
We acknowledge useful discussions with  Francesc Ferrer, Chris Mihos, and Harsh Mathur. This work was supported in part by the U.S. DOE under Contract No. DE-FG02-91ER40628 and the NSF under Grant No. PHY-0855580.

\end{document}